\documentclass[11pt,a4paper]{article}
\usepackage[T1]{fontenc}
\usepackage[utf8]{inputenc}
\usepackage{authblk}

\usepackage{geometry}
\usepackage{amsfonts,amsmath,amssymb,amsthm,nicefrac}
\usepackage{ulem}
\usepackage{cancel}
\usepackage{yhmath}
\usepackage{arcs}
\usepackage{graphicx}
\usepackage{url}
\usepackage{subcaption}
\title{\bf The global indicator of  classicality of an arbitrary $\boldsymbol{N}$\--level quantum system}
\author[1,2,3]{Vahagn Abgaryan}
\author[1,4,5]{Arsen Khvedelidze}
\author[1]{Astghik Torosyan}
\affil[1]{Laboratory of Information Technologies, Joint Institute for Nuclear Research, Dubna, Russia}
\affil[2]{A.I. Alikhanyan National Science Laboratory (Yerevan Physics Institute), Yerevan, Armenia}
\affil[3]{Department of Mathematics, Czech Technical University, Prague, Czech Republic}
\affil[4]{A. Razmadze Mathematical Institute, Iv. Javakhishvili Tbilisi State University, Tbilisi, Georgia}
\affil[5]{Institute of Quantum Physics and Engineering Technologies, Georgian Technical University, Tbilisi, Georgia}

\date{ }

\begin{document}

\maketitle

\begin{abstract}
 It is commonly accepted that a deviation of the Wigner quasiprobability distribution of a quantum state  from a  proper statistical distribution signifies its nonclassicality.  
 Following this ideology,  we introduce the global indicator $\mathcal{Q}_N$ for quantification of ``classicality\--quantumness'' correspondence in the form of the functional on the orbit space $\mathcal{O}[\mathfrak{P}_N]$ of the $SU(N)$ group adjoint action on the state space $\mathfrak{P}_N$
 of an $N$\--dimensional quantum system.
 The  indicator $\mathcal{Q}_{N}$ is defined as a relative volume of a subspace $\mathcal{O}[\mathfrak{P}^{(+)}_N] \subset \mathcal{O}[\mathfrak{P}_N]\,,$ where the 
 Wigner quasiprobability distribution is positive. An algebraic structure of $\mathcal{O}[\mathfrak{P}^{(+)}_N]$  is revealed and exemplified by a single qubit $(N=2)$ and single qutrit $(N=3)$. For the Hilbert-Schmidt ensemble of qutrits the dependence of global indicator on the moduli parameter of the Wigner quasiprobability distribution has been found.
\end{abstract} 

\tableofcontents

\newpage 

\section{Introduction}

Over the past decades, a number of  witnesses  and measures of  nonclassicality of quantum systems have been formulated (see e.g. \cite{Lemos2018,KZ,Mandel1979}).
Most of them are based on the primary impossibility of a classical statistical description of quantum systems. Particularly, the non-existence of positive definite  probability distributions serves as a certain indication 
of nonclassicality of a physical system.
\footnote{Furthermore, the negativity of quasiprobability distributions has been shown to be a resource for quantum computation \cite{Veitch,Albarelli2018}.} 

In the present note, we will focus on the problem of quantifying the nonclassicality of quantum systems associated with a finite-dimensional Hilbert space by studying the non-positivity of the Wigner quasiprobability distributions (the Wigner function, or shortly WF) \cite{Wigner1997,FerrieMorrisEmerson,ConnellWigner1981,ConnellWigner1984}. 
Our treatment  is based on  the recent publications~\cite{AV,AVA}, where the Wigner quasiprobability distribution $W^{(\boldsymbol{\nu})}_\varrho(\Omega_N)$ of an $N\--$level quantum system is constructed via the dual pairing, \begin{equation}
\label{eq:WignerFunction}
W^{(\boldsymbol{\nu})}_\varrho(\Omega_N) = \mbox{tr}\left[\varrho
\,\Delta(\Omega_N\,|\,\boldsymbol{\nu})\right]\,,
\end{equation} 
of the density matrix $\varrho$ \--- an element of a quantum state space $\mathfrak{P}_N$: 
\begin{equation}\label{eq:StateSpace}
    \mathfrak{P}_N =\{ X \in M_N(\mathbb{C}) \ |\ X=X^\dagger\,,\quad  X \geq 0\,,  \quad \mbox{tr}\left( X \right) = 1   \}\,,
\end{equation}
and an element of dual space $\Delta(\Omega_N\,|\,\boldsymbol{\nu}) \in \mathfrak{P}^\ast_N\,$\---  the  so-called Stratonovich-Weyl (SW) kernel.  
The dual space $\mathfrak{P}^\ast_N$ is defined as:
\footnote{The algebraic equations in (\ref{eq:SWspace})
define a family of $s\--$parametric SW kernels. Further, in the text, the  $s$-dimensional moduli parameter  $\boldsymbol{\nu}=(\nu_1, \nu_2, \dots, \nu_s ) , \,s \leq N-2\, $  (see details in \cite{AVA})
will be used to distinguish the corresponding  Wigner distributions (\ref{eq:WignerFunction}).
}
\begin{equation}
\label{eq:SWspace}
    \mathfrak{P}^\ast_N=\{ X \in M_N(\mathbb{C}) \ |\ X=X^\dagger\,,\quad \mbox{tr}\left( X \right) = 1\,, 
    \quad
   \mbox{tr}\left( X^2 \right) = N 
    \}\,,
\end{equation}
and SW  kernel is a mapping between phase space $\Omega_N$  and dual space $\mathfrak{P}^\ast_N\,.$
Assuming that SW kernel $\Delta(\Omega_N)\,$ has the  isotropy group $H\in U(N)$ of the form 
\[
H={U(k_1)\times U(k_2) \times U(k_{s+1})}\,,
\]
we identify phase-space $\Omega_N$\, as a complex flag manifold,  
$$
\Omega_N \,\to\, \mathbb{F}^N_{d_1,d_2, \dots, d_s}=
{U(N)} / {H}\,, 
$$
where 
$(d_1, d_2, \dots, d_s)$ is a sequence of positive integers with sum
$N $, such that  $k_1=d_1$ and $k_{i+1}=d_{i+1}-d_i$ with $d_{s+1}=N\,.$

The Wigner function  defined in eqs. (\ref{eq:WignerFunction}) - (\ref{eq:SWspace}) possesses all the properties of a proper statistical  distribution except for the non-negativity of the latter. From a physical point of view, the positiveness of probability distributions is a fundamental element of the classical statistical paradigm. Therefore, if WF attains negative values, it is undeniable that a physical system shows some  ``nonclassical'' behaviour.  
Following this observation,  we introduce the \textit{global indicator of classicality} $\mathcal{Q}_N$ characterizing the degree of closeness of a quasiprobability distribution to a proper one.  Commonly used measures of deviation from classicality are defined as functionals either on a quantum state space (the measures based  on the distance from the base ``classical state''),  or on phase space (the measures which depend on the volume of a phase space region where WF is negative \cite{KZ}). In contrast to this approach, we follow an alternative one, the so-called ``minimal description'' when characteristics of quantum systems are given exceptionally in the terms of $SU(N)\--$invariants. In other words, we intend to define the global indicator $\mathcal{Q}_N$ as a functional  over the unitary orbit space $\mathcal{O}[\mathfrak{P}_N]$. 
With this aim, let us introduce:
\begin{enumerate}
\item[]{\bf Definition 1}
The unitary orbit space $\mathcal{O}[\mathfrak{P}_N]$ is the quotient space under the equivalence relation imposed by the adjoint $SU(N)$  action on the state space  $\mathfrak{P}_N $ with quotient (canonical) mapping:
\begin{equation}
\label{eq:coset}
\pi:\  \mathfrak{P}_N \longrightarrow   \mathcal{O}[\mathfrak{P}_N] ={\mathfrak{P}_N }/{SU(N)} \,;
\end{equation}
\item[]{\bf Definition 2 } The subset $\Omega^{(+)}_N[\varrho] $ of phase space $\Omega_N\,,$ where the Wigner function of a given state $\varrho$ is non-negative, is 
\begin{equation}
\label{eq:Omega+}
\Omega^{(+)}_N[\varrho] =
\{\, x \in \Omega_N\,\ |\ W_\varrho(\Omega_N) \geq 0\, \,\}
\,;  
\end{equation}
\item[]{\bf Definition 3 } The subspace 
$\mathfrak{P}_N^{(+)} \subset  \mathfrak{P}_N$  is composed from states $\varrho$ so that
\begin{equation}
\label{eq:P+}
   \mathfrak{P}^{(+)}_N = \{\varrho \in \mathfrak{P}_N\,\ | \ 
\Omega^{(+)}_N[\varrho] = \Omega_N\, \ \};
\end{equation}
\item[]{\bf Definition 4 } The subset $\mathcal{O}[\mathfrak{P}^{(+)}_N]$ represents the image of $\mathfrak{P}_N^{(+)}$ under the quotient mapping (\ref{eq:coset}):
\begin{eqnarray}
\label{eq:OritP+}
\mathcal{O}[\mathfrak{P}^{(+)}_N]= \pi[\mathfrak{P}_N^{(+)}]= \{\pi(x)\, \ | \ x \in \mathfrak{P}_N^{(+)}\,\}\,. 
\end{eqnarray}
\end{enumerate}

Using the definitions above, we introduce the 
\textit{global indicator of nonclassicality}  $\mathcal{Q}_N $ of an N-dimensional quantum system as  the following  ratio:
\begin{equation}
\label{eq:indQ}
\mathcal{Q}_N =
\frac{\mbox{Volume of orbit subspace}\  \mathcal{O}[\mathfrak{P}_N^{(+)}]}{
    \mbox{Volume of orbit space}\  \mathcal{O}[\mathfrak{P}_N]}\,.
\end{equation}
In order to make this definition self-consistent, we assume that:
\begin{itemize}
\item{$\mathcal{O}[\mathfrak{P}_N]\,, \Omega^{(+)}_N[\varrho]\,, 
\mathfrak{P}_N^{(+)}]\,$  and $\mathcal{O}[\mathfrak{P}_N^{(+)}]$ are 
open, connected sets of $\mathbb{R}^{n}$\,;}
\footnote{In favor of this assumption, note that 
WF  is certainly non-negative for any state 
the Bloch vector of which lies inside  the ball of radius $r_*(N) = \sqrt{N+1}/{(N^2-1)}\,.$
}

\item{The volume of the orbit space in  (\ref{eq:indQ}) is associated with a measure induced by the quotient mapping $\pi$ from certain  Riemannian metric on  $\mathfrak{P}_N$\,.}\footnote{In the next section, the global indicator will be computed with respect to the metric corresponding to the Hilbert-Schmidt distance between density  matrices \cite{ZS2003}.}
\end{itemize}
In order to perform efficient  computations of $\mathcal{Q}_N$\,,  it is necessary  to have, instead of  implicit definitions (\ref{eq:P+}) and (\ref{eq:OritP+}),  a  more constructive representation  of the space 
$\mathcal{O}[\mathfrak{P}_N^{(+)}]\,.$
With this aim we remind some facts on the stratified  structure  of state space $\mathfrak{P}_N\,.$ 
First of all, note that  $U(N)$  automorphism of the  Hilbert space of an $N\--$level   quantum system induces the adjoint $SU(N)$ action  on 
the state space: 
\begin{equation}
\label{eq:UP}
    g\cdot \varrho = g \varrho g^\dagger\,, \qquad  g\in SU(N)\,.
\end{equation}
The group action (\ref{eq:UP}) sets an equivalence relations  between elements of $\mathfrak{P}_N\,$ and gives rise to $SU(N)$ orbit classification. Formally,  a subgroup $H_x \subset SU(N)$ is the  \textit{isotropy group (stabilizer)} of a point $x\in \mathfrak{P}_N\,,$
\[
H_x =\{g\in SU(N)\ |
\ g\cdot x =x\}\,,
\]
and points $x,y \in \mathfrak{P}_N $ are said to be of the same type if their stabilizers $H_x$ and $H_y$ are conjugate subgroups of $SU(N)$ group. The \textit{orbit type of the point} $x \in \mathfrak{P}_N$ is given by the conjugacy class of the corresponding isotropy group $[H_{x}]\,.$ 
Up to conjugation in $SU(N)$\,, the isotropy groups $H_x$ are in one-to-one correspondence with the Young diagrams corresponding to a possible decomposition of $N$ into non-negative integers. Hence,  for given $N$ for any $[H_\alpha]\,, \ \alpha = 1, 2, \dots, P(N)$ one can associate the \textit{stratum}   $\mathfrak{P}_{[H_\alpha]}\,,$  defined as the set of all points of $\mathfrak{P}_N$
whose stabilizer is conjugate to $H_\alpha$:
\footnote{The  strata ${\mathfrak{P}}_{(H_\alpha)}$  are determined by this set of equations and inequalities.}
\begin{equation}
   \mathfrak{P}_{[H_\alpha]}: =\big\{
x \in \ \mathfrak{P}_N|\  
 H_x \mbox{~is~conjugate~to}\  H_\alpha\big\} \,.
\end{equation}
The union of sets  $\mathfrak{P}_{[H_\alpha]}$ gives  the decomposition of state space $\mathfrak{P}_{N}$ into orbit types
\begin{equation}
\mathfrak{P}_N=\bigcup_{\mbox{orbit types}}{\mathfrak{P}}_{[H_\alpha]}\,.
\end{equation}
Having in mind the above notions and argumentation, we can formulate the following assertion.

{$\bullet$\,\underline{{\bf Proposition I}} $\bullet$\,}
\textit{Let $\boldsymbol{r}^\downarrow =\{r_1, r_2, \dots, r_N\}$ and $ \boldsymbol{\pi}^\uparrow =\{ \pi_N, \pi_{N-1}, \dots , \pi_1\}$
be eigenvalues of a density matrix $\varrho$ and SW kernel $\Delta(\Omega_N\,|\,\boldsymbol{\nu})\,,$ arranged  in decreasing and increasing orders respectively. Then,}
\begin{itemize}
\item[(i)]\textit{The Wigner function $W_\varrho(\boldsymbol{\theta})$ of any state $\varrho \in \mathfrak{P}_N$ is bounded and there exist $\boldsymbol{\theta}_{-}\,, \boldsymbol{\theta}_{+} \in \Omega_N$ such that 
\[
W_\varrho(\boldsymbol{\theta}_{-})=
\inf_{\boldsymbol{\theta} \in \Omega_N}\,   W_\varrho(\boldsymbol{\theta})
\,,
\qquad 
W_\varrho(\boldsymbol{\theta}_{+})=
\sup_{\boldsymbol{\theta} \in \Omega_N}\, W_\varrho(\boldsymbol{\theta})\,;
\]
}
\item[(ii)]If  $\varrho_1, \varrho_2 \in \mathfrak{P}_{[H_\alpha]}\,,$ then  extreme values of the corresponding Wigner functions
are related as follows,  
\begin{equation}
\inf_{\boldsymbol{\theta}}\,   W_{\varrho_1}(\boldsymbol{\theta})=
\inf_{\boldsymbol{\theta}}\,   W_{\varrho_2}(\boldsymbol{\theta})
\,,
\qquad 
\sup_{\boldsymbol{\theta}}\, W_{\varrho_1}(\boldsymbol{\theta})=
\sup_{\boldsymbol{\theta}}\, W_{\varrho_2}(\boldsymbol{\theta})\,;
\end{equation}
\item[(iii)]\textit{$\mathcal{O}[\mathfrak{P}_N^{(+)}]\,$  can be identified as  a dual cone  of a subset 
$\mathcal{O}[\mathfrak{P}_N] \subset \mathbb{R}^{N-1}$:}
\begin{equation}
\label{eq:OP+}
\mathcal{O}[\mathfrak{P}_N^{(+)}] =  \
\left\{\, \boldsymbol{\pi} \in \mathcal{O}[\mathfrak{P}^\ast_N]\ \, |\ \,  ( \boldsymbol{r}^\downarrow, \boldsymbol{\pi}^\uparrow) \geq 0,  \quad \forall\, \boldsymbol{r} \in \mathcal{O}[\mathfrak{P}_N]\, \right\}\,,
\end{equation}
\textit{where the dual pairing $(\,,\, )$ in (\ref{eq:OP+}) is }
\begin{equation}
\label{eq:dualpair}
( \boldsymbol{r}^\downarrow, \boldsymbol{\pi}^\uparrow) =
r_1\pi_{N}+r_2\pi_{N-1} +\dots +r_{N}\pi_1\,.
\end{equation}
\end{itemize}
The correctness of the above proposition stems from the following  observations.
At first, according to our construction, an $N\--$level system is associated 
with a symplectic manifold $\Omega_N\,,$ which is compact. Secondly,  the  Wigner
distributions of trace-class operators are continuous functions (cf. discussion in \cite{Daubechies1982}. Hence,  in accordance to the 
multivariable Weierstrass extreme value theorem, the Wigner function 
attains its extreme values on the $\Omega_N$.
Moreover, the absolute maximum and minimum must occur at a critical point of  WF in $\Omega_N$ or at a boundary point of $\Omega_N$.
Some technical details of the proof of Proposition  I are given in the Appendix.

The article is organized as follows. The next section is  devoted to a brief exposition of necessary facts about  WF of finite-dimensional systems mainly borrowed from our recent articles \cite{AV,AVA}. 
In  Section \ref{sec:Brkhofpolytope}.
we present a reinterpretation  of the Wigner distributions as a functions defined over  the space of the unistochastic matrices and  describe their  continuation  to the whole Birkhoff polytope. 
With the aid of this extension,  the global extrema of WF is derived. In Section \ref{sec:Bounds}.,   using the lower and upper bounds for WF the orbit subspace $\mathcal{O}[\mathfrak{P}^{(+)}_N]\,, $ the  global $\mathcal{Q}\--$indicators for $N=2$ (qubit) and $N=3$ are obtained.
Final remarks are collected in Section \ref{sec:Summary}.

\section{Basic settings}
\label{sec:WFNlevel}

\noindent{$\bullet$ \bf Wigner function of 
$N$\--level system $\bullet$\,}
A  density matrix $\varrho$ and Stratonovich-Weyl kernel $\Delta(\Omega_N\,|\,\boldsymbol{\nu})$ 
obey the following decompositions into  the Lie algebra $\mathfrak{su}(N)$ and its dual
$\mathfrak{su}(N)^*$,
\begin{eqnarray}
\label{eq:DMsuN}
\varrho &=& \frac{1}{N}\mathbb{I}_N + \frac{1}{N}\,\imath\, \mathfrak{su}(N)\,,\\
\Delta(\Omega_N\,|\,\boldsymbol{\nu}) &=& \frac{1}{N}\mathbb{I}_N + \kappa\,\frac{1}{N}\,\imath\, \mathfrak{su}(N)^\ast \,,
\label{eq:SWsuN}
\end{eqnarray}
where $\kappa =\sqrt{{N(N^2-1)}/{2}}\,$ is a normalization constant.
It is convenient to use the orthonormal  Hermitian basis  $\boldsymbol{\lambda} = (\lambda_1, \lambda_2, \dots, \lambda_{N^2-1} ) $ of the
$\mathfrak{su}(N)$ algebra and rewrite 
the density matrix (\ref{eq:DMsuN}) in the Bloch form
\begin{equation}
\label{eq:rhoN}
\varrho_{\boldsymbol{\xi}}=\frac{1}{N}\left(I+\sqrt{\frac{N\left(N-1\right)}{2}}\left(\boldsymbol{\xi},\boldsymbol{\lambda}\right)\right)\,, 
\end{equation}
where  $\boldsymbol{\xi}\, $  stands for the $(N^2-1)$\--dimensional Bloch vector.
Parallel to (\ref{eq:rhoN}), we will extensively use the Singular Value Decomposition (SVD) of SW kernel,
\begin{equation}
    \label{eq:SVDSW}
\Delta(\Omega_N\,|\,\boldsymbol{\nu}) =\frac{1}{N}U(\Omega_N)\left( I +\kappa
    \sum_{\lambda_s \in \mathfrak{h}}\mu_s(\boldsymbol{\nu})
    \lambda_s\right)U^\dagger(\Omega_N)\,,
\end{equation}
where $\mathfrak{h}$ is the Cartan subalgebra $\mathfrak{h}\in \mathfrak{su}(N)\,.$ Under  these conventions, the algebraic  equations in  (\ref{eq:SWspace}) define the following family of the Wigner functions:
\begin{equation}
\label{eq:WFCartan}
W^{(\boldsymbol{\nu})}_{\boldsymbol{\xi}} (\theta_1,\theta_2, \dots,  \theta_d)=\frac{1}{N}\left[1 + \frac{N^2-1}{\sqrt{N+1}}\,(\boldsymbol{n}, \boldsymbol{\xi})\right]\,.
\end{equation}
In  (\ref{eq:WFCartan})  the Wigner function dependence on a point of phase space $\Omega_N$  with coordinates $(\theta_1, \theta_2,  \dots, \theta_d)$ 
\footnote{The number $d$ of independent variables $\theta $ in the Wigner function varies depending on the dimension of the isotropy group of SW kernel, $d=\mathrm{dim}_{\mathbb{C}}\, \mathbb{F}^N_{d_1,d_2, \dots, d_s} $\,.}
is encoded in the  $(N^2-1)$\--dimensional  vector  $\boldsymbol{n}$ 
given by the linear
superposition:
\begin{eqnarray}
\label{eq:nvector}
\boldsymbol{n} = \mu_3(\boldsymbol{\nu}) 
\boldsymbol{n}^{(3)} + 
\mu_8(\boldsymbol{\nu}) \boldsymbol{n}^{(8)} + \dots+ 
\mu_{N^{2}-1}(\boldsymbol{\nu})
\boldsymbol{n}^{(N^{2}-1)}\,. 
\end{eqnarray}
The real coefficients $\mu_3(\boldsymbol{\nu}), \mu_8(\boldsymbol{\nu}), \dots, \mu_{N^2-1}(\boldsymbol{\nu})$ characterize a family of the Wigner functions through their  dependence on coordinates $\boldsymbol{\nu}$ of  the  moduli  space,  $\mathcal{P}_N(\boldsymbol{\nu})\,.$ The moduli space $\mathcal{P}_N(\boldsymbol{\nu})\,$ represents a  spherical  polyhedron on a unit sphere, which is in one-to-one correspondence with  an ordering of the eigenvalues of SW kernel\footnote{Detailed description of the moduli space $\mathcal{P}_N(\boldsymbol{\nu})\,$ is presented in \cite{AVA}.}
\begin{equation}
\label{eq:Nsphere}
    \mu_3^2(\boldsymbol{\nu}) + \mu_8^2(\boldsymbol{\nu})
    +\dots+\mu^2_{N^{2}-1}(\boldsymbol{\nu})=1\,.
\end{equation}
The orthonormal vectors     $\boldsymbol{n}^{(3)}, \boldsymbol{n}^{(8)}, \dots, 
\boldsymbol{n}^{(N^{2}-1)}$ in 
(\ref{eq:nvector})
are specified by  $N-1$ basis elements
$\lambda_3, \lambda_8, \dots, \lambda_{N^2-1},$
of the Cartan subalgebra $\mathfrak{h} \subset \mathfrak{su}(N)$:
\begin{equation}
\nonumber \boldsymbol{n}^{(s^2-1)}_\mu = \frac{1}{2}\,\mbox{tr}\left( U\lambda_{s^2-1}U^\dagger\lambda_\mu \right)\,.
\end{equation}
Finally, it is worth to mention that the Wigner function (\ref{eq:WFCartan}) is a normalized distribution,  
\begin{equation}
 \int_{\Omega_N} \mathrm{d}\Omega_N\, W_\varrho(\Omega_N)  = 1\,,
\end{equation}
with the measure $\mathrm{d}\Omega_N $ determined from the normalized Haar measure $\mathrm{d}\mu_{SU(N)}$ on the $SU(N)$ group manifold: 
\[
\,{\mathrm{d}\mu_{SU(N)}} =
\frac{1}{N \mbox{Vol}(H)}\,
\mathrm{d}\Omega_N \times \mathrm{d}\mu(H)\,.
\]
Here, 
$\mbox{Vol}(H)$ is the volume of the isotropy group of SW kernel computed with respect to the measure $\mathrm{d}\mu(H)$ which is induced by the corresponding  embedding of  $H$ into $SU(N)$\,.
 
\noindent{$\bullet$ \bf Orbit space of 
$N$\--level system $\bullet$\,} Similarly to 
(\ref{eq:SVDSW}), writing down the SVD of a density matrix $\varrho$ with fixed, say decreasing order of eigenvalues 
$\boldsymbol{r}=(r_1, r_2, \dots, r_N)$\,,
\begin{equation}
\label{eq:SVDdensity matrix}
\varrho = U
\begin{pmatrix}
r_1 & \cdots & 0 \\
\vdots& \ddots & \vdots \\
0 & \cdots & r_N
\end{pmatrix}
U^\dagger\,,
\end{equation}
we realise the quotient  mapping (\ref{eq:coset}) 
from the state space $\mathfrak{P}_N$ to the orbit space $\mathcal{O}[\mathfrak{P}_N]$   
in the form of 
ordered $(N-1)$\--simplex:
\begin{equation}\label{eq:orderedsymplex}
C_{N-1} = \{\  \boldsymbol{r} \in \mathbb{R}^N\,\ \biggl|\,\ 
\sum_{i=1}^{N} r_i = 1, \quad 1\geq r_1\geq r_2 \geq \dots \geq r_{N-1}\geq r_N \geq 0 \ \}\,.
\end{equation}
In the present note we mainly focus on the Wigner functions (\ref{eq:WFCartan}) of qubit  $(N=2)$ and qutrit $(N=3)$ and thus will deal  with a 1-simplex, a line segment, and  a 2-simplex, a triangle, correspondingly.

\section{The Wigner distribution as a function on the Birkhoff polytope}
\label{sec:Brkhofpolytope}

In this section we rewrite the Wigner distribution in the form of a function on the so-called Birkhoff polytope $\mathcal{B}_N$\, \cite{Bengtsson2005Comun}.  
The Birkhoff polytope $\mathcal{B}_N\,$  is the polytope of the  \textit{bistochastic or doubly stochastic $N\times N$ complex matrices} obeying the following conditions:
\[ B_{ij} \geq 0\,, \quad \sum_{i=1}^N B_{ij}=1\,, \quad \sum_{j=1}^N B_{ij}=1\,.\]
Precisely speaking, the Wigner function of an $N\--$level system is defined over the subset of bistochastic matrices called \textit{unistochastic}. If the matrix $B$ is expressible via a unitary matrix $U$:
\[B_{ij}= \mid U_{ij}\mid^2 \,, \qquad \forall \, i,j=1,2,\dots N\,,
\]
then it is unistochastic.
The following proposition establishes this relation.

{$\bullet$\,\underline{\bf Proposition II}\, $\bullet$}\textit{ 
Let us assign to a matrix $B \in \mathcal{B}_N$ the bilinear form on $\mathbb{R}_{+}^N$: 
\begin{equation}
\label{eq:Bill}
    (\boldsymbol{x}, \boldsymbol{y})_{B} = \left(\boldsymbol{x}\,,  B\boldsymbol{y} \right) = \sum_{ij}B_{ij} x_i y_j \,.
\end{equation}
Then the  Wigner quasiprobability distribution of an  $N\--$level system can be identified as the bilinear form (\ref{eq:Bill})
with matrix $B$ from a subset 
$\, \mathcal{U}_N \subset \mathcal{B}_N $ of a unistochastic matrices:
\footnote{Note that 
for $ N\geq 3$ the set of unistochastic matrices is not convex.}}
\begin{equation}
 \label{eq:WFBirkhof}
 W_\varrho(\Omega_N) = \left(\boldsymbol{r}^{\downarrow}\,, \boldsymbol{\pi}^{\downarrow}\right)_{B}\biggl|_{B=|U|^2}\,,
 \end{equation}
\textit{evaluated at ordered vectors $\boldsymbol{r}^{\downarrow}$ and $\boldsymbol{\pi}^{\downarrow}$ whose components are eigenvalues  of a density matrix $\varrho$ and SW kernel  $\Delta$,  respectively. 
}

Based on the {Proposition II},  we are  able to 
study  the problem of determination of a global extrema  of WF as follows.  Noting that an analogous problem 
for the bilinear form $\left(. \,, .\right)_B$ is well studied,  we define the \textit{continuation of the Wigner distribution} as a function $W(B)\,,$ whose domain of definition is the whole Birkhoff polytope
\begin{equation}
\label{eq:WB}
W(B) := \left(\boldsymbol{r}^{\downarrow}\,, \boldsymbol{\pi}^{\downarrow}\right)_{B}\,.
\end{equation}
Applying  the  Birkhoff–von Neumann theorem to the function  $W(B)$\,,  one can determine its global maximum and minimum. Next step is to analyze  the fate of the extrema after a restriction of (\ref{eq:WB}) to the subspace  of unistochastic matrices.
The following conjecture aims to answer this question.

{$\bullet$\,{\bf Proposition III}\, $\bullet$}\textit{
The Wigner  quasiprobability distribution  function defined on a set of unistochastic matrices attains  the global maximum  $W^{(+)}$ and global minimum $W^{(-)}$ at permutation matrices 
\begin{eqnarray}
\label{eq:PminPmax}
&&P_{\min}=
\begin{pmatrix}
0 & \cdots & 1 \\
\vdots& 1 & \vdots \\
1 & \cdots & 0
\end{pmatrix}
\,, \qquad  P_{\max}=
 \begin{pmatrix}
1 & \cdots & 0 \\
\vdots& 1 & \vdots \\
0 & \cdots & 1
\end{pmatrix} \,,
\end{eqnarray}
with the following values 
\begin{eqnarray}
W^{(-)}&= &  \lim_{B\to P_{\min}} W = 
\left(\boldsymbol{r}^{\uparrow}\,, \boldsymbol{\pi}^{\downarrow}\right)
\,,\\
     W^{(+)} &=&\lim_{B\to P_{\max}} W =\left(\boldsymbol{r}^{\downarrow}\,, \boldsymbol{\pi}^{\downarrow}\right)\,.
\end{eqnarray}
}
For a formal discussion of this conjecture we refer  to the Appendix,  while  here we  only give two argumentations in favor of this conjecture.  
The  first one is the Birkhoff–von Neumann theorem \cite[p.~36]{Bhatia} according to which 
$\mathcal{B}$ is the convex hull of all $ N\times N$ permutation matrices.  There is at least one decomposition of  $\mathcal{B}$:
\begin{equation}
 \mathcal{B}_N = \sum_{i}^k \kappa_i P_i\,, \qquad \sum_{i}\kappa_i =1\,, \quad \kappa_i \geq  0\,, 
\end{equation}
with $k \leq (n-1)^2 +1$ permutation matrices $P_i\,,$
corresponding to the vertices of the Birkhoff polytope.
Due to this theorem, the bilinear form 
$\left(. \,, .\right)_B$
assumes  its extremum for the set of extreme points consisting of the permutations (\ref{eq:PminPmax}) mentioned in the conjecture,
\begin{eqnarray}
\min_B\left(x\,, y\right)_{B} &=& \left(x\,, y\right)_{P_{\min}} = \sum_{i}x_i^{\uparrow}y_i^{\downarrow}\,,\\
\max_B\left(x\,, y\right)_{B} &=& \left(x\,, y\right)_{P_{\max}} = \sum_{i}x_i^{\downarrow}y_i^{\downarrow}\,.
\end{eqnarray}
The second argumentation in favor of the conjecture is that the space of unistochastic matrices contains all permutation matrices  and $P_{\min}$
and $P_{\max}$ are among them. 

Therefore, for a given SW kernel with the eigenvalues 
$\boldsymbol{\pi}^{\downarrow}= \{ \pi_1, \pi_2, \dots, \pi_N \}$ and a density matrix with spectrum 
$\boldsymbol{r}^{\downarrow}= \{ r_1, r_2, \dots, r_N \}$
the knowledge of the global minimum of WF  provides us information on the subset  $\mathcal{O}[\mathfrak{P}^{(+)}_N]$\, from the  inequality  $W^{(-)} \geq 0\,:$ 
\begin{eqnarray}
\label{eq:O+}
   \mathcal{O}[\mathfrak{P}^{(+)}_N]:\quad
   \{ \boldsymbol{r} \in C_{N-1}\ 
   | \ \left(
   \boldsymbol{r}^{\uparrow}\,, \boldsymbol{\pi}^{\downarrow}\right) \geq 0\,\} \,.
\end{eqnarray}

Based on these results, in the next section we explicitly evaluate the rate  of quantumness-classicality for low-dimensional systems, such as a qubit and a qutrit.

\section{Global indicator of nonclassicality of qubit and qutrit}
\label{sec:Bounds}

Summarising discussions of the previous section,  the Wigner function satisfies the following inequality:
\begin{equation}
\label{eq:LUBounds}
   W^{(-)}_N \leq  W(\Omega_N) \leq W^{(+)}_N\,, 
\end{equation}
where 
\begin{equation}
\label{eq:BoundsN}
W^{(-)}_N= \sum_{i=1}^{N}\pi_ir_{N-i+1}\,,
    \qquad W^{(+)}_N =\sum_{i=1}^{N}\pi_ir_i\,.
\end{equation}
Below considering  the inequalities (\ref{eq:LUBounds}) for two low cases, $N=2$ and $N=3$\,, we will obtain 
an explicit  parameterization of subspaces $\mathcal{O}[\mathfrak{P}_2^{(+)}]$ and $\mathcal{O}[\mathfrak{P}_3^{(+)}]$  of the orbit  space corresponding to a positive WF of a single qubit and single qutrit. 

\noindent{$\bullet $ {\bf Positivity of the lower bound $W^{(-)}_2$ }$\bullet $\,} For a simplest $N=2$ level system, a single qubit, the  density matrix expanded over the Pauli 
$\boldsymbol{\sigma}$\--matrices  is characterised by a 3-dimensional Bloch vector $\boldsymbol{\xi}=(\xi_1, \xi_2, \xi_3)$: 
\begin{equation}
    \varrho = \frac{1}{2}\left(I + (\boldsymbol{\xi}, \boldsymbol{\sigma})
    \right)\,.
\end{equation}
According to (\ref{eq:SVDSW}), the spectrum of SW kernel for a qubit  is unique, and 
assuming the decreasing ordering of eigenvalues it is 
\begin{equation}
\mbox{spec}\left(\Delta_2\right) =\biggl\{ \frac{1+\sqrt{3}}{2}\,, \frac{1-\sqrt{3}}{2} \biggl\}\,.
\end{equation}
Taking into account the above expressions, the lower and upper  bounds (\ref{eq:BoundsN}) for a qubit are:  
\begin{equation}
   W^{(\mp)}_2 = \frac{1}{2} \mp 
   \frac{\sqrt{3}}{2}\,{\vert\boldsymbol{\xi}\vert}\,. 
\end{equation}
Therefore, the Wigner function of a qubit is positive definite inside the Bloch ball of radius  $r_*(2) < {1}/\sqrt{3}\,. $

\noindent $\bullet$ {\bf $\mathcal{Q}$-indicator of a single qubit\,} $\bullet$ Based on the above derived constraint on a qubit states with nonnegative WF, the global indicator $\mathcal{Q}$ of quantumness can be evaluated after fixation of the measure on  the orbit space of a qubit $\mathcal{O}[\mathfrak{P}_2]$.
The measure $\mathrm{d}\mu_{{}_\mathrm{H-S}}$ on $\mathfrak{P}_2$ associated with the Hilbert-Schmidt ensemble of qubits has a product form 
\begin{equation}
\label{eq:HSmeasure}
  \mathrm{d}\mu_{{}_\mathrm{H-S}}= (r_1-r_2)^2\, \mathrm{d}r_1\wedge \mathrm{d}r_2 \times \mathrm{d}\mu_{\frac{\mathrm{SU(2)}}{\mathrm{U(1)}}}\,,
\end{equation}
where  
$\mathrm{d}\mu_{\frac{\mathrm{SU(2)}}{\mathrm{U(1)}}}$ is the measure 
on the coset $SU(2)/U(1)\,,$ induced from the normalized Haar measure on $SU(2)$ group. The factor in (\ref{eq:HSmeasure}) which depends on 1-simplex coordinates  
$r_1$ and  $r_2$ defines the measure on the orbit space $\mathcal{O}[\mathfrak{P}_2]\,.$
Thus,  computation of the indicator  $\mathcal{Q}$ of a qubit reduces to evaluation of  the ratio of two simple integrals,  
\begin{equation}
    \mathcal{Q}_2 = \frac{\mbox{Vol}\left( \mathcal{O}[\mathfrak{P}^{(+)}_2]\right)}{\mbox{Vol}
    \left(\mathcal{O}[\mathfrak{P}_2]\right)}
    \,=\,
    \frac{\displaystyle{\int_{0} ^{\frac{1}{\sqrt{3}}} r^2 dr}}{\displaystyle{
    \int_{0}^{1} r^2 dr}} = \frac{1}{3\sqrt{3}}=
    0.19245\,.
\end{equation}

\noindent{$\bullet $ {\bf Positivity of the lower bound $W^{(-)}_3$ }$\bullet $\,} 
For further study introduce  two types of coordinates  on the orbit space of a qutrit.  The first parameterization takes  into account the algebraic structure of a density matrix of a qutrit states:
\begin{eqnarray}
\label{eq:specrho}
r_1=\frac{1}{3}+\frac{1}{\sqrt{3}}\xi_3+\frac{1}{3}\xi_8\,,
\quad
r_2=\frac{1}{3}-\frac{1}{\sqrt{3}}\xi_3+\frac{1}{3}\xi_8\,, \quad 
r_3=\frac{1}{3}-\frac{2}{3}\, \xi_8\,.
\end{eqnarray}
In terms of $\xi_3$ and $\xi_8$ the ordered 2-simplex  is mapped to the domain  $\mathcal{O}[\mathfrak{P}_3]$ defined by the following set of inequalities:  
\begin{equation}
\label{eq:Orbitxi3xi8}
\mathcal{O}[\mathfrak{P}_3]\, \colon \ \biggl\{ \xi_3, \xi_8 \in \mathbb{R}\,\ \biggl|\,\ 0\leq \xi_3 \leq\frac{\sqrt{3}}{2}\,, \quad \frac{\xi_3}{\sqrt{3}} \leq \xi_8 \leq \frac{1}{2}\biggl\} \,.  \end{equation}
The second useful set of coordinates,
$(r, \varphi)\,,$
on  the orbit space of a qutrit  is  given  by the following map:
\begin{equation}
\label{eq:mapsimplexMaclauren}
 \xi_3=\sqrt{3}r\sin\left(\frac{\varphi}{3}\right)\,, 
 \qquad 
\xi_8=\sqrt{3}r\cos\left(\frac{\varphi}{3}\right)\,, 
\qquad 0 \leq \varphi \leq \pi\,.
\end{equation}
Under  the transformation (\ref{eq:mapsimplexMaclauren}) the ordered 2-simplex of a qutrit is mapped into the domain on upper half-plane with coordinates $x=r\cos\varphi,\, y=r\sin\varphi\,,$ outlined by the  trisectrix of Maclaurin (see the grey region depicted in Fig.\ref{Fig:MaclaurinTrisectrix3}):
\begin{figure}[h]
\centering
\includegraphics[scale=0.4]{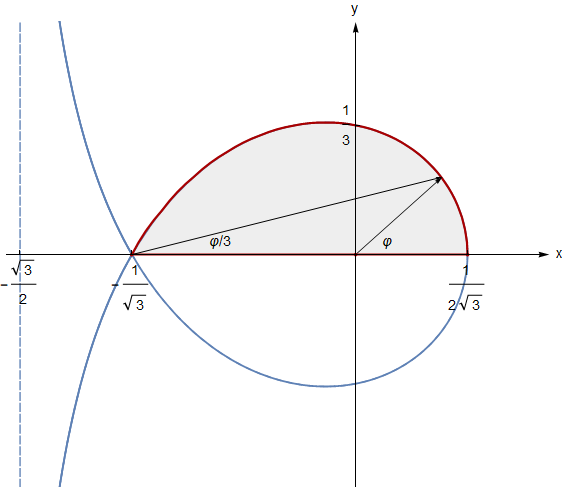}
\caption{The trisectrix of Maclaurin intersecting x-axis at two points,  $(\frac{1}{2\sqrt{3}},0)$ and $(-\frac{1}{\sqrt{3}},0)$. On $(x,y)$ plane the equation of  this  curve in polar coordinates
    $x=r\cos\varphi,\, y=r\sin\varphi\,,$ 
    reads: $r(\varphi, \frac{1}{\sqrt{3}} )=\frac{1}{2\sqrt{3} \cos({\varphi}/{3})}$\,. The orbit space of a qutrit $\mathcal{O}[\mathfrak{P}_3]$ is given by the domain in grey.}
\label{Fig:MaclaurinTrisectrix3}
\end{figure}
\begin{equation}\label{eq:OrbitQutrit}
 \mathcal{O}[\mathfrak{P}_3]\,: \  
 \biggl\{\, r \geq 0\,,  \varphi \in [0, \pi]\,\ \biggl|\, \ \cos\left(\frac{\varphi}{3}\right) \leq \frac{1}{2\sqrt{3}r} \, \biggl\}\,.
\end{equation}

According to the analysis given in \cite{AVA}, the algebraic equations (\ref{eq:SWspace}) for the  eigenvalues of SW kernel of a qutrit  have one-parametric solution which can be written as
\begin{eqnarray}
\label{eq:specDelta}
\pi_1=\frac{1}{3}+\frac{2}{\sqrt{3}}
\mu_3+\frac{2}{3}\,\mu_8,\quad
\pi_2=\frac{1}{3}-\frac{2}{\sqrt{3}}\mu_3+\frac{2}{3}\,\mu_8, \quad 
\pi_3=\frac{1}{3}-\frac{4}{3}\,\mu_8\,.
\end{eqnarray}
Here the parameters $\mu_3$ and $\mu_8$ are Cartesian coordinates of a segment of a unit circle with the apex angle $\zeta$:  
\begin{equation}
\label{eq:muzeta}
\mu_3=\sin\zeta\,, \qquad \mu_8=\cos\zeta\,, 
\qquad 0 \leq \zeta \leq \frac{\pi}{3}\,.
\end{equation}
 It is worth to note that the apex angle $\zeta$ determines the value of a 3-rd order polynomial SU(3)-invariant  of SW kernel $\Delta(\Omega_3) | \nu)$:
\[
\cos(3\zeta)=-\frac{27}{16}\det\left(\Delta(\Omega_3 | \nu )\right)-
\frac{11}{16}\,
\]
with the moduli parameter $\nu$:
\begin{equation}
\nu=\frac{1}{3}-\frac{4}{3}\cos(\zeta)\,, \quad \zeta \in [0,\  {\pi}/{3}]\,.
\end{equation}

\begin{figure}[h!]
  \centering
  \begin{subfigure}[b]{0.4\linewidth}
    \includegraphics[width=\linewidth]{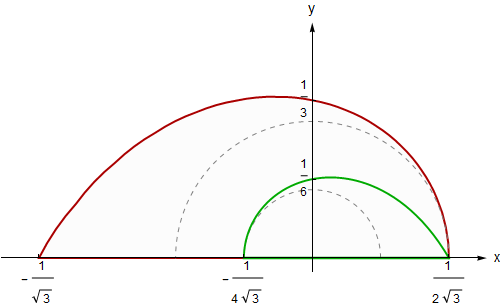}
     \caption{$\zeta=0$}
  \end{subfigure}
  \begin{subfigure}[b]{0.4\linewidth}
    \includegraphics[width=\linewidth]{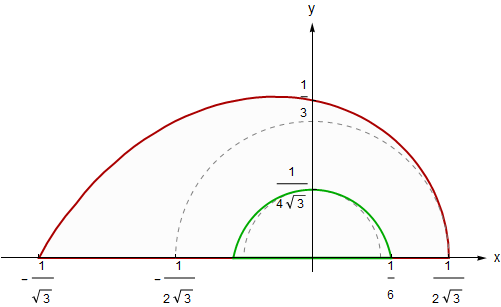}
    \caption{$\zeta=\frac{\pi}{6}$}
  \end{subfigure}
  \begin{subfigure}[b]{0.4\linewidth}
    \includegraphics[width=\linewidth]{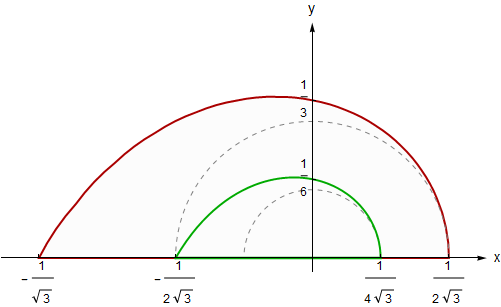}
    \caption{$\zeta=\frac{\pi}{3}$}
  \end{subfigure}
  \caption{
  The state space of a qutrit divided into bands. 
  The Wigner function is always positive (necessarily has some negative values) inside (outside) the region enclosed by the dashed inner (outer) semicircle regardless of the choice of the kernel. Inside the region enclosed by the kernel-dependent inner solid curve  the Wigner function is always positive for the specific choice of the kernel.
  }
  \label{Fig:PositivityQutritLBounds}
\end{figure}

Having these ingredients for a  density matrix (\ref{eq:specrho})  and  SW kernel (\ref{eq:specDelta}),  the 
straightforward evaluation of (\ref{eq:BoundsN}) for $N=3$ gives  
\begin{eqnarray}
\label{eq:LowerB}
    W_3^{(-)}&=& \frac{1}{3}-\frac{4r}{\sqrt{3}}\cos\left( \zeta + \frac{\varphi}{3} -\frac{\pi}{3}\right)\,,
\\
    W_3^{(+)}&=& \frac{1}{3}+\frac{4r}{\sqrt{3}}\cos\left( \zeta - \frac{\varphi}{3}\right)\,.
\label{eq:UpperB}
\end{eqnarray}
From the expression  (\ref{eq:LowerB}) it follows 
that a subspace of the orbit space $\mathcal{O}[\mathfrak{P}^{(+)}_3]$\, where WF is positive  reads: 
\begin{eqnarray}\label{eq:PosiDomain}
&& \mathcal{O}[\mathfrak{P}^{(+)}_3]\, : \biggl\{\, r \geq 0\,,  \varphi \in [0, \pi]\,\ \biggl|\, \ \cos\left(\frac{\varphi}{3} +\zeta -\frac{\pi}{3}\right) \leq \frac{1}{4\sqrt{3}r}\,\biggl\}\,.
\end{eqnarray}
Comparing (\ref{eq:PosiDomain}) with the qutrit orbit space (\ref{eq:OrbitQutrit}) 
we conclude that $\mathcal{O}[\mathfrak{P}^{(+)}_3]$\, lies inside the qutrit orbit space $\mathcal{O}[\mathfrak{P}_3] $ as it is 
shown in Fig. \ref{Fig:PositivityQutritLBounds}.
Here it is in order to make few comments on a shape of  $\mathcal{O}[\mathfrak{P}^{(+)}_3]$\,:
\begin{itemize}
\item For $0 \leq r\leq \frac{1}{4\sqrt{3}}$ the lower bound $W_3^{(-)}$ is positive for all $\zeta$ and $\varphi$;
\item For $\frac{1}{2\sqrt{3}}\leq  r \leq \frac{1}{\sqrt{3}}$ the lower bound $W_3^{(-)}$  is always negative;
\item For intermediate values $\frac{1}{4\sqrt{3}} \leq  r \leq \frac{1}{2\sqrt{3}}$ the lower bound 
$W_3^{(-)}$  becomes negative only for certain values of $\zeta $ and $\varphi$.
\end{itemize}

\noindent{$\bullet $ {\bf $\mathcal{Q}$-indicator of a single qutrit \,}$\bullet $\,} 
The global indicator  of nonclassicality of a qutrit is given by the ratio of volumes
\begin{equation}
\label{eq:Q3}
    \mathcal{Q}_3 = \frac{\mbox{Vol}\left( \mathcal{O}[\mathfrak{P}^{(+)}_3]\right)}{\mbox{Vol}
    \left(\mathcal{O}[\mathfrak{P}_3]\right)}
    \,.
\end{equation}
To evaluate these  volume integrals we need to specify a measure on the orbit space $\mathcal{O}[\mathfrak{P}_3]$.
Similarly to a qubit case, we assume that a qutrit state space $\mathfrak{P}_3$ is endowed with the Hilbert-Schmidt metric:  
\begin{equation}
\label{eq:metric}
 \mathrm{g} = 4\, \mathrm{tr}\left(\mathrm{d}\varrho\otimes\mathrm{d}\varrho
 \right)\,.
\end{equation}
In terms of the Bloch coordinates $\boldsymbol{\xi} = (\xi_1, \xi_2, \dots , \xi_8) $ of a qutrit, 
\begin{equation}
    \varrho= \frac{1}{3}\left(\mathbb{I}+ \sqrt{3}\,(\boldsymbol{\lambda},\boldsymbol{\xi})\right)\,,
\end{equation}
the metric (\ref{eq:metric}) gives the standard Euclidean volume form on $\mathfrak{P}_3$: 
\begin{equation}
\label{eq:dvol}
 \omega =\left(\frac{8}{3}\right)^{4}\, \mathrm{d}\xi_{1}\wedge\mathrm{d}\xi_{2}\wedge\cdots\wedge
 \mathrm{d}\xi_{8}\,.
\end{equation}
Now in order to compute the corresponding induced form on the orbit space $\mathcal{O}[\mathfrak{P}_3]$\,, we rewrite (\ref{eq:dvol}) 
in terms of the SVD of the density matrix 
\begin{equation}\label{eq:SVDdensity}
  \varrho = UDU^{\dag}\,.  
\end{equation}
Since the measure of a singular and degenerate matrices is zero, we consider a generic spectrum 
 $D=\mbox{diag}||r_1, r_2, r_3||$\, with descending order of eigenvalues
$1 > r_1 > r_2 > r_3 > 0\,.$ This means that  
the arbitrariness of $U$ is given by the torus $T$ of $SU(3)$ and the volume form (\ref{eq:dvol})  is useful to write down in an adaptive SVD coordinates, 
\begin{equation}\label{eq:measureigen}
\omega= (r_1-r_2)^2(r_1-r_3)^2(r_2-r_3)^2\,\mathrm{d}r_1\wedge\mathrm{d}r_2\wedge\mathrm{d}r_3\wedge\omega_{{}_\mathrm{SU(3)/T}}\,.
\end{equation}
For illustrative reasons it is convenient to pass  from 2-simplex Catrtesian coordinates $r_1, r_2, r_3$ to  the polar variables $r$ and $\varphi\,,$ 
introduced in (\ref{eq:mapsimplexMaclauren}).
As a result, the 
volume form (\ref{eq:measureigen}) on the orbits space $\mathcal{O}[\mathfrak{P}_3]$ reduces to the following expression:
\begin{equation}
\label{eq:qutritmeasure}
  \omega_{\mathcal{O}[\mathfrak{P}_3]}  =
 r^7 \sin^2{\varphi}\,
 \mathrm{d}r\wedge\mathrm{d}\varphi\,.
\end{equation}
Computing the volume integrals in (\ref{eq:Q3}) with respect to the measure (\ref{eq:qutritmeasure}) over the orbit space of a qutrit (\ref{eq:OrbitQutrit}) and its subspace were WF is positive, we find  an explicit dependence of the global indicator of classicality on the SW kernel moduli parameter $\zeta\,$:
\begin{eqnarray}
\label{eq:Ratio}
\mathcal{Q}_3 (\zeta)= \displaystyle{
\frac{\int_0^\pi d\varphi\int_{0} ^{\frac{1}{4\sqrt{3} \cos{\left(\frac{\varphi}{3}+\zeta-\frac{\pi}{3}\right)}}} r^7 \sin^2(\varphi) dr}{
\int_0^\pi d\varphi\int_{0} ^{\frac{1}{2\sqrt{3} \cos{\frac{\varphi}{3}}}} r^7 \sin^2(\varphi) dr }}=\frac{1}{128}\,\frac{1+20 \cos^2{\left(\zeta -{\pi }/{6}\right)}}{ \left(-1+4\cos^2{\left(\zeta -{\pi }/{6}\right)}\right)^5}\,.
\end{eqnarray}

\begin{figure}[h!]
\centering
\includegraphics[scale=0.4]{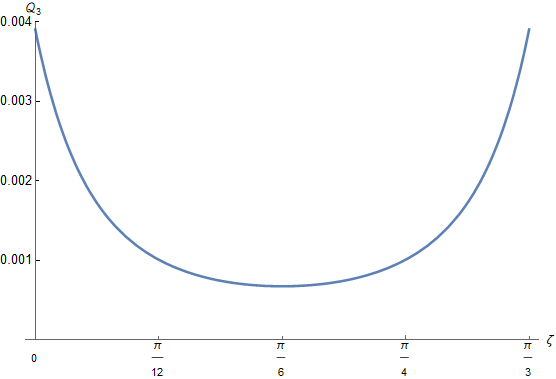}
\caption{$\mathcal{Q}\--$indicator as  a function of SW kernel moduli parameter $\zeta\,$ for the  Hilbert-Schmidt qutrit.}
\label{Fig:QLowerBound}
\end{figure}
The straightforward  calculations show that the indicator $\mathcal{Q}_3(\zeta)$ attains 
the absolute minimum 
at a qutrit modili parameter 
$\zeta = {\pi}/{6}\,,$
\[
\min_{\zeta \in [0,\frac{\pi}{3}]}\, \mathcal{Q}_3(\zeta)= \mathcal{Q}_3\left(\frac{\pi}{6}\right) =
 \frac{7}{2^7\, 3^4} \approx 0.000675\,,
\]
corresponding to SW kernel with the spectrum:  
\begin{equation}
\mbox{spec}\left(\Delta_3\right)=\biggl|\biggl|\,\frac{1+2\sqrt{3}}{3}\,, \frac{1}{3}\,, \frac{1-2\sqrt{3}}{3}\,\biggl|\biggl|\,.
\end{equation}
In Fig.\ref{Fig:QLowerBound} the dependence of 
$\mathcal{Q}_3$ on the moduli parameter $\zeta $ is 
shown.

\section{Summary}
\label{sec:Summary}

In the present article we introduce the global indicator of classicality  of quantum $N\--$dimensional system. This indicator directly measures the portion of its unitary orbit space  which is associated to states 
admitting conventional statistical interpretation in terms of a true probability distributions. During the study an interesting  relation between the properties of the Wigner quasiprobability distributions and structure of the Birkhoff polytopes has been found out.  
It seems that this  relation deserves attention, and in our future  publication we will come back to the problem of a classical-quantum correspondence from this point of view.

\section*{Appendix}
\label{sec:Appendix}

In this Appendix we discuss the global extrema problem of a function over the  unitary orbits of a Hermitian matrix. 

\noindent{\bf Problem} \textit{Let $A$ be a positive definite Hermitian matrix and $B$ be a Hermitian matrix. Consider the adjoint unitary orbits,    $\mathcal{O}_B= gBg^\dagger\,$,  with $ g \in SU(N)$. Find the  global extrema of the function }
 \begin{equation}
  \label{eq:FunctOrbit} 
   \Phi(g)= \mathrm{tr}(AgBg^\dagger)\,.
   \end{equation}
To find the extrema of (\ref{eq:FunctOrbit}), one can apply a standard method of calculus used for a problem of determination critical points of functions. To be accurate, consider matrices $A$ and $B$  whose spectrum is of the  following form: 
\begin{eqnarray}
\boldsymbol{\mu}^{\downarrow}(A)&=&\{\mu_1(A) \overbrace{(1, \dots, 1)}^{k_1(A)}\,;\, \mu_2(A)\overbrace{(1, \dots, 1)}^{k_2(A)}\,;\, \dots \,;\, \mu_s(A)\overbrace{(1, \dots, 1)}^{k_s(A)}\}\,,
\\
\boldsymbol{\mu}^{\downarrow}(B)&=&\{\mu_1(B) \overbrace{(1, \dots, 1)}^{k_1(B)}\,;\, \mu_2(B)\overbrace{(1, \dots, 1)}^{k_2(B)}\,;\, \dots \,;\, \mu_s(B)\overbrace{(1, \dots, 1)}^{k_s(B)}\}\,.
\end{eqnarray}
The elements of spectra of both matrices are  arranged  in the decreasing order:  
\begin{equation}
\mu_1(A) > \mu_2(A) > \dots  > \mu_s(A) 
    \qquad   
\mbox{and} 
    \qquad 
\mu_1(B) > \mu_2(B) > \dots  > \mu_s(B)\,.
\end{equation}
The degrees of degeneracy $(k(A), k(B))$ of matrices $A$ and $B$ are  constrained  by the relations,  
 $ \sum_{i=1}^s k_i(A) =r_A $  and 
 $ \sum_{i=1}^s k_i(B) =r_B\,.$ The SVD decompositions for matrices 
\begin{equation}
\label{eq:SVDAB}
A=V D_A V^\dagger\,, \qquad 
B=W D_B W^\dagger\,
\end{equation}
are not unique and a family of unitary matrices $V$ and $W$ in (\ref{eq:SVDAB}) can be built as follows. 
Let us denote by $V^{\downarrow}$ the unitary matrix constructed of the right eigenvectors 
of matrix $A$, disposed in the  correspondence with the decreasing order of its  eigenvalues. Then the  most general family of unitary matrices  diagonalizing  $A$ reads 
\begin{equation}
\label{eq:Generic}
V = V^{\downarrow}
\begin{pmatrix}
V_1 & \cdots & 0 \\
\vdots& \ddots & \vdots \\
0 & \cdots & V_s
\end{pmatrix}
P\,,
\end{equation}
where $V_1,  \dots , V_s $
are arbitrary unitary matrices of  order $k_1,  \dots, k_s$  respectively and $P$ is the transposition matrix 
\[
P =|| \boldsymbol{e}_{i_1}\,, \boldsymbol{e}_{i_2}\,, \dots , \boldsymbol{e}_{i_N}||,
\]
with $N\--$dimensional vectors $\boldsymbol{e}_{j}$ having everywhere zeros except 1 in the  $j\--$place. 
The right multiplication,  $ AP$\,,
transposes the columns $j \to i_j\,,  j=1, \dots N\,.$
Below the same construction for the unitary matrix $W$ will be used as well. 

Straightforward computations show that the necessary condition of  extrema for $\Phi(g)$ can be written as 
\begin{equation}
  \label{eq:NCCextrem}
  \mathrm{d}\Phi(g)= \mbox{tr}\left([O_B, A] w_g \right)=0\,,
\end{equation}
where
\begin{equation}
w_g =\mathrm{d}g g^\dagger= \frac{\imath}{2}\,\sum_{a,i =1}^{N^2-1}(w_g)^a_i\lambda_a \mathrm{d}\vartheta^i\,
\end{equation}
is the Maurer-Cartan 1-form on $SU(N)\,$ group.
The equation 
(\ref{eq:NCCextrem}) tells us that extrema of $\Phi(g)$ are  realized for all points of  the orbits  $\mathcal{O}_B=g_c Bg_c^\dagger\,, $  commuting with $A$\,: \footnote{
The condition (\ref{eq:NCCextrem}) represents a system of linear homogeneous  equations $(w_g)^a_i x_a = 0\,$ with unknown $x_a$  and apart from the trivial solution, $x_a=0\,,$ can have other solutions corresponding to singular points occuring at $\det||(w_g)^a_i|| = 0$. Recalling that $\det||w_g|| = \sqrt{\det||\mathrm{g}_{{}_\mathrm{U(N)}}||}\,,$  and the explicit expression for the Haar measure $\sqrt{\det||\mathrm{g}_{{}_\mathrm{U(N)}}||}\,\mathrm{d}\vartheta_1\cdots  \mathrm{d}\vartheta_N$ in terms of eigenvalues  of $U(N)$ element  
$$
\sqrt{\det||\mathrm{g}_{{}_\mathrm{U(N)}}||}=\frac{1}{(2\pi)^NN!}\prod_{1\leq i<j \leq n }\biggl|e^{i\vartheta_i} - e^{i\vartheta_j} \biggl|^2\,,$$ 
we associate a set of singular solutions to  (\ref{eq:NCCextrem}) to  a variety of possible types of degeneracies of the eigenvalues of the unitary matrices,  $\vartheta_{i_1}=
\vartheta_{i_2}=\dots =\vartheta_{i_k} $.  
  }  
\begin{equation}
    \label{eq:condofextremum}
    \ [A,\mathcal{O}_B] = 0\,.
\end{equation}
This equation has a solution $g_c = VW^\dagger$  with 
the unitary matrices $V$ and $W$ diagonalizing $A$ and $B$  respectively.
According to (\ref{eq:Generic}), the matrices $V$ and $W$ constitute  a family of  diagonalising unitary matrices. One can see that a set  of corresponding critical points $g=g_c$ of  $\Phi(g)\,$ is discrete. As a result of (\ref{eq:Generic}), 
for given $\mbox{spec}(A)$ and $\mbox{spec}(B)$ the  extrema are determined by permutations $P$
\[
\Phi(g)\biggl|_{g=g_c}=
\mbox{tr}(D_AD_B)= \mbox{tr}\left(\boldsymbol{\mu}^{\downarrow}(A)P^T\boldsymbol{\mu}^{\downarrow}(B)P \right) \,.
\]
Among these exrtema, the minimum and maximum are identified using the well-known result of majorisation of two vectors $x,y \in \mathbb{R}^N$ (cf. \cite[p.~49]{Bhatia}):  
\begin{equation}
\langle x^{\downarrow}, y^{\uparrow} \rangle \leq 
\langle x, y \rangle 
\leq 
\langle x^{\downarrow}, y^{\downarrow} \rangle\,.
\end{equation}
Hence, finally,  global extrema of $\Phi(g)$ read, 
\begin{eqnarray}
  \min_{g\in g_c} \Phi(g) &=& \mbox{tr}\left(\boldsymbol{\mu}^{\downarrow}(A)\, \boldsymbol{\mu}^{\uparrow}(B)\right) \,,\\
\max_{g\in g_c} \Phi(g) &=& \mbox{tr}\left(\boldsymbol{\mu}^{\downarrow}(A)\, \boldsymbol{\mu}^{\downarrow}(B)\right) \,.
\end{eqnarray}


\end{document}